\documentstyle[referee]{l-aa}

\begin{document}
\thesaurus{4(10.19.3; 10.06.2; 13.09,1)}
\title{The radial scale length of the Milky Way}

\author{C. Porcel
\inst{1}
\and F. Garz\'on
\inst{2}
\and J. Jim\'enez-Vicente
\inst{1}
\and E. Battaner
\inst{1}}

\institute{Departamento de F\'{\i}sica Te\'orica y del Cosmos.
        Universidad de Granada. Spain 
\and Instituto de Astrof\'{\i}sica de Canarias. E-38200, La Laguna,
Tenerife
Spain} 
\maketitle
\begin{abstract}
        The radial scale length of the exponential component of the disc of
the Milky Way has been determined in the near infrared. We have used
the TMGS (Two Micron Galactic Survey) database which contains
positions and K-magnitudes of about 700.000 stars distributed in
serveral regions along the galactic plane. From those we have selected
areas more than $5^\circ$ off the plane to minimize the effect of
extinction and the contributions of the young disc; and with
longitudes ranging from $30^\circ$ to $70^\circ$, to avoid
contaminations from the central bulge, the bar and the molecular ring,
in the inner end, and from the local arm, the warp and the truncation
of the disc, in the outer end. We observed stars with magnitude 
$ 9 \leq m_K \leq 10 $. The use of the NIR K-band also reduces 
the effect of extinction. In
the observation region, $ m_K=9.5 ~\mbox{ mag}$ stars are K2-K5 III 
stars, with an absolute magnitude that is nearly constant, which also greatly
simplifies the problem. We have obtained the value $2.1 \pm 0.3~ \mbox{ kpc}$
for the radial scale length, which is a typical value when compared with 
external galaxies of similar types.

\end{abstract}
\keywords{Galaxy: structure -- Galaxy: fundamental parameters --
Infrared: galaxies}

\section{Introduction}

        It is well known that normal discs of spiral galaxies have a 
component which can be fitted to an exponential (Freeman, 1970) with
a radial scale length $h$. There are two basic reasons to undertake a
new determination of the parameter $h$ in the Milky Way:

        a) There is a very large scatter in the values found in the
literature. For instance, Kent, Dame and Fanzio (1991) after reviewing
previously reported values, found that they range from 1 to 5.5 kpc, and
proposed an intermediate value of 3 kpc. van der Kruit (1986) obtained
5.5 kpc; Lewis and Freeman (1989) gave a value of 2.5 kpc, and so did
Robin, Crez\'e and Mohan (1992), among others. This uncertainty is also  
too large in the NIR. Mikami and Ishida (1981) obtained a value of 1 kpc,
Jones et al. (1981) 2 kpc, and Eaton et al. (1984) 3 kpc. Values of $h$ in
the NIR seem to be lower, but even this fact remains uncertain, despite
its interest in the dynamic evolution of the disc.

        b) Even if $h$ is simply a fitting parameter for a distribution
with frequent deviations from an exponential, it has became a natural
radial length unit, very useful when comparing spiral galaxies with 
different sizes or unknown distances, and physically more significant
than true length of angle units. The position of any morphological feature
or typical lengths of disc phenomena are usually and hopefully expressed
taking $h$ as unit. Some of these features in our own galaxy may be
considered normal or exceptional depending on the value of this 
normalizing length.

        Probably the large scatter of data is a result of managing a large
amount of data concerning different regions of the sky, which are difficult
to analyse because of the contribution of many other complex components,
such as bulge, star formation ring associated with the gas ring, bar,
spiral arms, warp and so on. However, a proper choice of the wavelength 
and the region of the sky may render this problem a very simple one, which
implies a more confident obtention of the parameter $h$

        This search specifically deals with, and aims at an improved
determination of the radial scale length of the Milky Way disc.

        We have chosen the K-band, in which extinction is very
        low. This also implies that the observed distribution closely
        resembles the true one. In 
addition, we have taken measurements at $b=5^\circ$, rendering extinction
negligible. We then surveyed a galactic longitude range from
$30^\circ$ to $70^\circ$. This practically eliminates the contribution
of bulge, ring and bar (Hammersley et al., 1994) which take place for 
$l < 30^\circ$, and the contribution of the Local Arm, warp and truncation
(Porcel et al., 1997), which take place for $l > 70^\circ$. The 
contribution of the other spiral arms (not the Local Arm) is minimized
taking $b=5^\circ$, as they remain closer to the plane. Another advantage
of working in the K-band is that the observed stars in the 9-10 magnitude
range are practically only K2-K5 III stars, through the surveyed sky zone.
This fact was realised after taking into account the previous analyses of 
Ruelas-Mayorga (1991), Wainscoat et al. (1992), Calbet et al. (1995)
and our own initial calculations (Porcel, 1997). The light coming from
these stars
is 80 \% of the total light from all stars in this magnitude range. This
greatly simplifies the problem as the luminosity function can be
approximated by a Dirac's delta function.

        Observations were carried out with the Carlos S\'anchez 1.5 m telescope
at the Teide Observatory, as part of the Two Micron Galactic Survey (TMGS)
project (Garz\'on et al., 1993; Hammersley et al., 1994; 
Calbet et al., 1995). Interpolation techniques were used to obtain a regular 
mesh from the series of discrete scans (Porcel, 1997; Porcel et al. 1997).

\section{Results}

        As explained above we consider K2-K5 III stars. The luminosity 
function $\phi$ is defined as usual, i. e. $\phi (M)dM$ gives the number
of stars per $\mbox{ pc}^3$ with an absolute K magnitude in the interval
$[M, M+dM]$. It is here approximated by a Dirac's function, i. e. 
$\phi (M)=\delta (M-M_0)$ centred at $M_0=-2.5 \mbox{mag}$. Its
true dispersion is only 0.6 mag. The effect of a non-vanishing
dispersion on absolute magnitude could be important, as the resulting
value for the scale length could be affected by some sort of Malmquist
bias. We have performed several numerical calculations using a
gaussian distribution for the luminosity function, obtaining similar
results. The final value of the estimated error is not affected
either, as it mainly arises from the fitting process. Even though
these kind of effects are difficult to evaluate, the approximation of
the luminosity function by a $\delta-{\mathrm function}$ was found to
be a satisfactory assumption. In the fundamental equation of star
counts (e. g. Mihalas and Binney, 1981; Gilmore and Reid, 1983; Calbet
et al., 1995) the density function is also taken into account. It
was assumed that:
\begin{equation}
n(R,z)=e^{-\frac{R-R_\odot}{h}-\frac{|z|}{h_z}}
\label{eq1}
\end{equation}
where $n(R,z)$ is the density function, i. e. the number density of stars
taking its local value as unit, $R$ is the galactocentric distance,
and $z$ is the
vertical coordinate. We take $R_\odot=8.5~\mbox{kpc}$. The vertical 
exponential profile and the
value of $h_z=200 \mbox{ pc}$ were adopted from Wainscoat et
al. (1992). We have repeated these calculations for different values
of $h_z$ and we have obtained similar results, i. e. neither the proposed
value of $h$ nor its estimated error are affected. Actually $h$ is found to be
noticeably independent of $h_z$. We therefore think that this choice
for $h_z$ cannot be an important source of errors.

        Defining as usual ${\cal A}(l,b)$, as the number of stars per 
squared degree with magnitude between 9 and 10, the fundamental equation
of star counts becomes
\begin{equation}
{\cal A}=\omega \int_9^{10}{dm} \int_0^{\infty} {C \delta (M-M_0)
e^{-\frac{R-R_\odot}{h}-\frac{|z|}{h_z}}r^2 dr}
\label{eq2}
\end{equation}
where $\omega$ is the solid angle, $C$ is a constant and $r$ is the distance
from the Sun. This expression is easily transformed into
\begin{equation}
{\cal A}(l)=C_3 e^{-\frac{R_0-R_\odot}{h}}
\label{eq3}
\end{equation}
where
\begin{equation}
R_0=(R_{\odot}^2+r_0^2 \cos ^2 b -2R_{\odot}r_0 \cos b \cos l)^{1/2}
\label{eq4}
\end{equation}
$C_3$ is another constant, and
\begin{equation}
r_0=10^{\frac{m-M_0+5}{5}}=2.5~ \mbox{ kpc}
\label{eq5}
\end{equation}

\begin{figure}
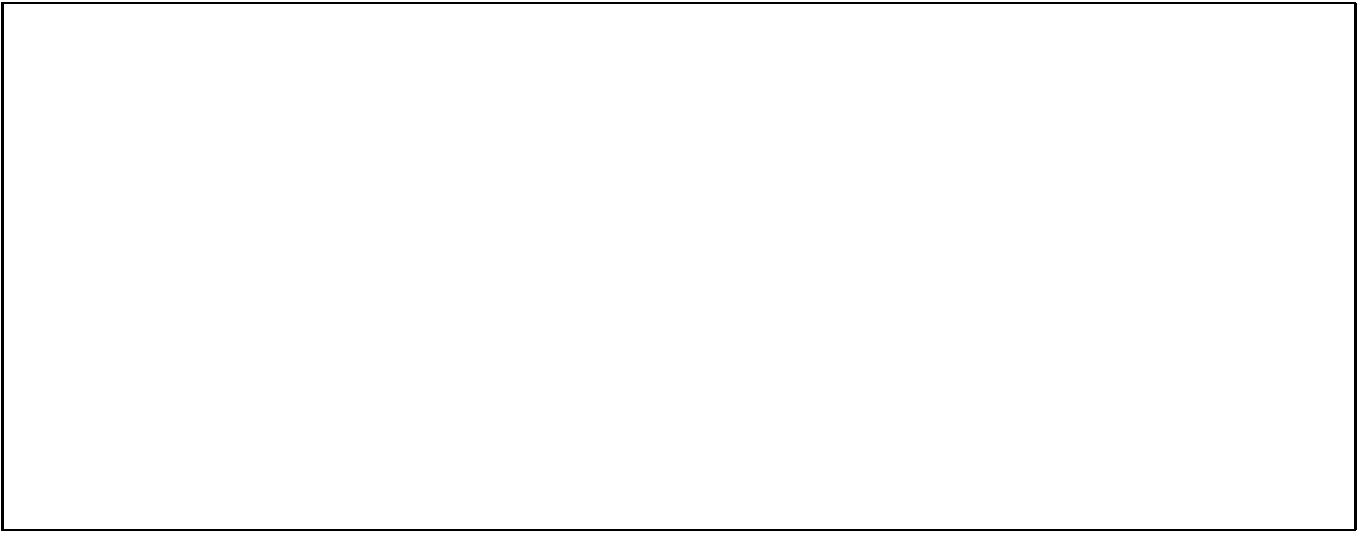

\picplace{7cm}
\caption{Log of the number of K=9-10 magnitude stars per squared degree as
a function of $R_0 (l)$ defined in the text.}
\label{fig1}
\end{figure}

        As $b$ is taken to be constant, $R_0$ is a function only of $l$, 
and therefore ${\cal A}$ is a function only of $l$. Equation
(\ref{eq3}) 
predicts a
linear function relation between $\log {\cal A}$ and $R_0$. The results are
plotted in Figure \ref{fig1}. In this figure, error bars represent
Poisson errors. After fitting (using a standard chi-squared method 
as described in Press et al., 1992), we obtained 
\begin{equation}
h=(2.1\pm 0.3)
~\mbox{ kpc}
\end{equation}

\section{Conclusions}

        The value we propose, $h=(2.1 \pm 0.3)~ \mbox{kpc}$ has been obtained
by choosing the most appropriate conditions for this task. This value is 
similar to typical values for other galaxies (de Grijs and van der
Kruit, 1996; Peletier et al., 1994; de Jong, 1996). This value seems
quite low when compared with the obtained values in the optical
(see for instance van der Kruit, 1986; and Barteldress and Dettmar, 
1994) but
it is well known than the exponential scale lengths decrease with
increasing wavelength (de Grijs and van der Kruit, 1996; Peletier et al.,
1994; de Jong 1996, Tully et al., 1996). 
Peletier et al. (1994)
found that the ratio $h_B/h_K$ is in the range 1.2-2 in a sample of 37
Sb's and Sc's
edge-on galaxies, with this ratio increasing with axis ratio, showing
that it is an effect that is mostly due to extinction. In this case 
our determination
of the radial scale length of the Milky Way is compatible with values
in optical bands as large as 4 kpc. Our value is virtually free of
extinction effects, so it is a proper determination of the true
scale length of the mass distribution of the Milky Way disc. A similar
value ($2.3 \pm 0.1$  kpc )has been obtained by Ruphy et al. (1996). Fux and Martinet
used indirect methods based on the asymmetric drift equation, 
and also obtained a similar value ($2.5^{+0.8}_{-0.6}$ kpc). We therefore
conclude that with respect to the radial scale length, our galaxy is
a typical one.


\begin{thebibliography}{}
\bibitem{Bart94} Barteldrees A. \& Dettmar J., 1994, A\&AS 103, 475-502
\bibitem{Calbet95} Calbet X., Mahoney T., Garz\'on F. \& Hammersley
P. L. 1995, MNRAS, 276, 301
\bibitem{deGrijs} de Grijs R. \& van der Kruit P. C. 1996, A\&AS 117, 19
\bibitem{deJong} de Jong R. S. 1996, A\&A 313, 45-64
\bibitem{Eaton84} Eaton N., Adams D. J. \& Giles A. B. 1984, MNRAS,
208, 41
\bibitem{Free70} Freeman K. C. 1970, ApJ, 160, 811
\bibitem{Fux94} Fux, R. \& Martinet L. 1994, A\&A 287, L21-L24
\bibitem{Garzon93} Garz\'on F., Hammersley P. L., Mahoney T., Calbet
X., Selby M. J., Hepburn I. D. 1993, MNRAS, 264, 773
\bibitem{Gilmore83} Gilmore G. \& Reid I. 1983, MNRAS, 202, 1025
\bibitem{Hammer94} Hammersley P. L., Garz\'on F. Mahoney T., Calbet X.
 1994, MNRAS, 269, 753
\bibitem{Jones81} Jones T. J., Asley M., Hyland M., Ruelas-Mayorga A.
 1981, MNRAS, 197, 413
\bibitem{Kent91} Kent S. M., Dame T. M. \& Fanzio G. 1991, ApJ, 378,
131
\bibitem{Lewis89} Lewis J. R. \& Freeman K. C. 1989, AJ 97, 139
\bibitem{Miha81} Mihalas D. \& Binney J. 1981, in {\em Galactic
Astronomy: Structure and Kinematics}, W. H. Freeman, San Francisco.
\bibitem{Mikami81} Mikami T. \& Ishida K. 1981, PASP, 33, 135
\bibitem{Pelet94} Peletier R. F., Valentijn E. A., Moorwood
A. F. M. \& Freudling W. 1994, A\&AS 108, 621
\bibitem{Porcel97a} Porcel C. 1997, Ph. D. Thesis, Univ. Granada
\bibitem{Porcel97b} Porcel C., Battaner, E. \& Jim\'enez-Vicente 
J. 1997, A\&A in press.
\bibitem{Press92} Press, W. H., Teukolsky S. A., Vetterling W. T. \& 
Flannery B. P. 1992, in {\em Numerical Recipes}, Cambridge Univ. Press 
\bibitem{Robin92a} Robin A. C., Creze M. \& Mohan V. 1992, A\&A, 265,
32
\bibitem{Ruelas91a} Ruelas-Mayorga R. A. 1991, Rev. Mex. Astrof. 22,
27-41
\bibitem{Ruphy96} Ruphy S., Robin A. C., Epchtein N., Copet E., Bertin
 E., Fouqu\'e P. \& Guglielmo F. 1996, A\&A 313, L21
\bibitem{Tully96} Tully, R. B., Verheijen, M. A. W., Pierce, M. J., 
Huang, J. S., Wainscoat, R. 1996, AJ 112, 2471
\bibitem{vdKruit86} van der Kruit P. C. 1986, A\&A, 157, 230
\bibitem{Wains92} Wainscoat R. J., Cohen M., Volk K. et al. 1992,
ApJS, 83, 111
\end{thebibliography}
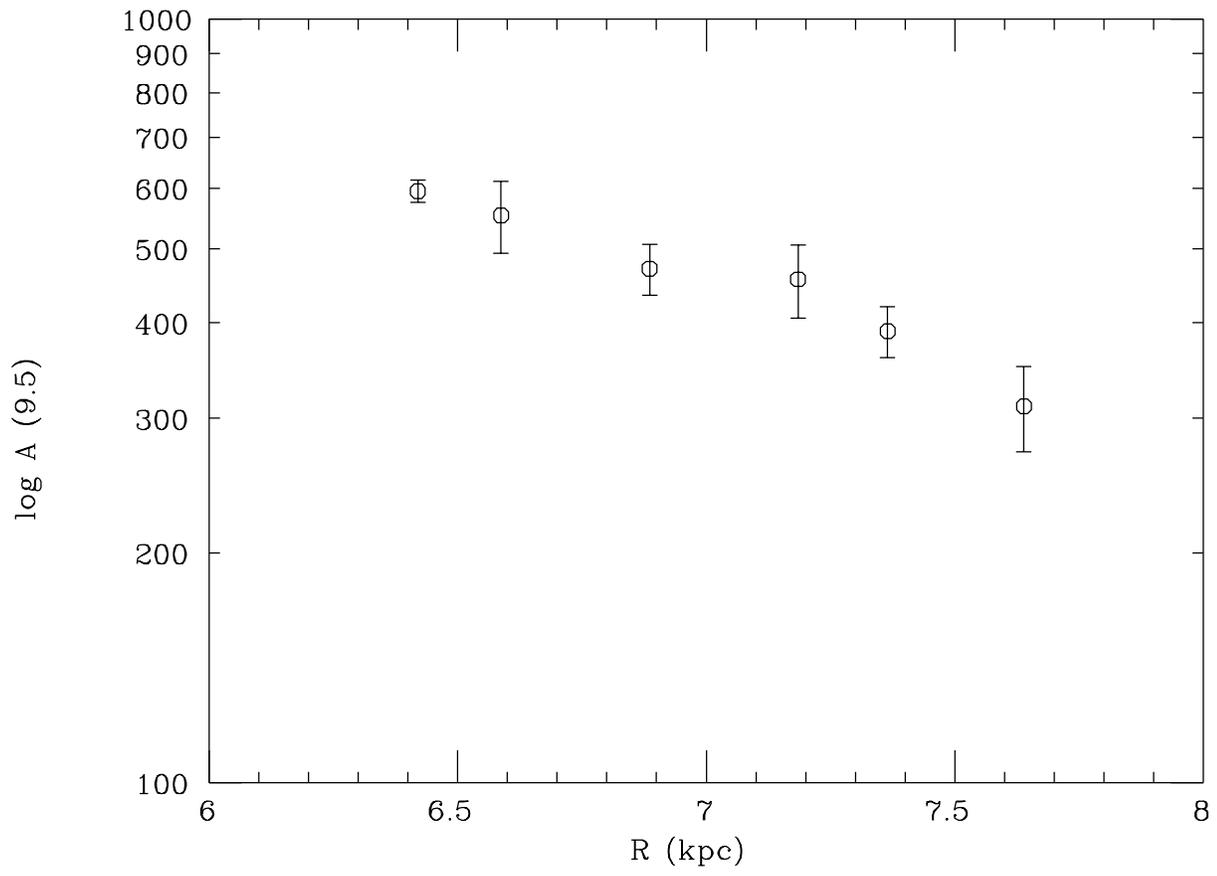
\end{document}